\documentclass[amsmath, amssymb,10pt, aps, prx, twocolumn, notitlepage, eqsecnum, longbibliography, superscriptaddress]{revtex4-1}
   
\usepackage{graphicx}
\usepackage{calc}
\usepackage{braket}
\usepackage{slashed}
\usepackage{wasysym}
\usepackage{dsfont}
\usepackage{amsthm,amsmath,amsfonts,amssymb,verbatim,color}
\usepackage{lipsum}
\usepackage{subfigure}
\usepackage{bm}
\usepackage{epsfig,slashed}
\usepackage{hyperref}
\usepackage{flafter}
\usepackage{booktabs}
\usepackage{makecell} 
\usepackage{bbm} 
\usepackage[export]{adjustbox} 
\usepackage{cancel}
\usepackage{epstopdf}
\usepackage{tikz}
\usepackage{array,color}
\usepackage{float}

\newcommand{\bA}{{\bf A}}
\newcommand{\bB}{{\bf B}}
\newcommand{\bE}{{\bf E}}

\newcommand{\bk}{{\bf k}}

\newcommand{\bq}{{\bf q}}

\newcommand{\br}{{\bf r}}
\newcommand{\bu}{{\bf u}}

\newcommand{\bzero}{{\bf 0}}

\def\brho{{\boldsymbol \rho}}

\newcommand{\hk}{{\hat k}}

\newcommand{\hn}{{\hat n}}

\newcommand{\hT}{{\hat T}}

\newcommand{\hpsi}{{\hat\psi}}

\newcommand{\cB}{{\cal B}}

\begin{document}
\title{Weyl hydrodynamics in a strong magnetic field}	

\author{Siyu Zhu}
\affiliation{Physics Department, University of California, Santa Cruz, California 95064, USA}

\author{Grigory Bednik}
\affiliation{Physics Department, University of California, Santa Cruz, California 95064, USA}

\author{Sergey Syzranov}
\affiliation{Physics Department, University of California, Santa Cruz, California 95064, USA}

\begin{abstract}
We study the hydrodynamic transport of electrons in a Weyl semimetal in a strong magnetic field.
Impurity scattering in a Weyl semimetal with two Weyl nodes is strongly anisotropic as a function of the 
direction of the field and is significantly suppressed if the field is perpendicular to the separation between the
nodes in momentum space.
This allows for convenient access to the hydrodynamic regime of transport, in which electron scattering is dominated by
interactions rather than by impurities. In a strong magnetic field, electrons move predominantly parallel to
the direction of the field,
and the flow of the electron liquid in a Weyl-semimetal junction resembles the Poiseuille flow of a liquid in a pipe.
We compute the viscosity of the Weyl liquid microscopically and find that it weakly depends on the magnetic field
and has the temperature dependence $\eta(T)\propto T^2$. The hydrodynamic flow 
of the Weyl liquid can be generated by a temperature gradient. 
The hydrodynamic regime in a Weyl-semimetal junction can be
probed via the thermal conductance $G_q(B,T)\propto B^2 T$ of the junction.
\end{abstract}


\maketitle


\section{Introducion}

Hydrodynamics has recently been receiving attention as a paradigm for describing transport in
sufficiently clean materials with strong electron correlations
not amenable to exact microscopic treatment.
Hydrodynamic description deals with macroscopic degrees of freedom,
such as the densities of particles and their momenta.

The hydrodynamic regime requires that the electron-electron scattering rate significantly exceed the electron-phonon  
and impurity scattering rates and is predicted to lead to such uncovnentional phenomena as 
Gurzhi effect~\cite{Gurzhi1,Gurzhi2} (growing conductance with increasing temperature),
current vortices~\cite{Levitov:vortices,Bandurin:grapheneHydro} and magnetic dynamos in electron liquids~\cite{Galitski:dynamo}.
Hydrodynamic transport is also often discussed 
as a possible mechanism behind the linear-in-$T$ resistivity in high-temperature superconductors~\cite{Davison:linearT,Hartnoll:linearT}.

Dirac materials in 2D (graphene~\cite{Narozhny:KarlsruheTeamGrapheneReview,Narozhny:grapheneReview,Lucas:grapheneReview}) and 3D (Weyl and Dirac semimetals~\cite{Sukhachov:hydroCollective,Gorbar:WeylHydro,Gorbar:hydroCS,Gorbar:hydroNonLocal,Galitski:dynamo,Sukhachov:convectionImpossible}) is another popular venue for theoretical studies of hydrodynamic effects.
Hydrodynamic flows in such systems simulate ultrarelativistic interacting matter and, in the case of two dimensions,
allow for convenient visualisation (see, e.g., Ref.~\cite{KuKim:grapheneFlowImaging}).

Despite extensive theoretical studies, achieving the hydrodynamic regime is rather challenging;
materials that allow for conclusive experimental observations of hydrodynamic transport are few and far between.
Such observations include manifestations of hydrodynamics in the nonlocal transport
in high-mobility $(Al,Ga)As$ heterostructures~\cite{Gupta:GaAsHydro,Braem:GaAsScanning},
magnetoresistive~\cite{Bockhorn:MRGaAs,Alekseev:GiantMRhydroExplanation,Mani:MRGaAs,Shi:MRGaAs} and
Gurzhi effects~\cite{DeJongMolenkamp:firstGurzhi,Gusev:hydroGaAs,Gusev:hydroHall} 
in $(Al,Ga)As$ heterostructures, deviations from the Wiedemann-Franz law~\cite{Jaoui:WiedemannFranz} in $WP_2$ and a combination of magnetotransport phenomena 
in $PdCoO_2$~\cite{Moll:PdCoO}. Graphene provides another popular playground for observing hydrodynamic phenomena~\cite{Bandurin:grapheneHydro,Bandurin:grapheneNegativeMR,Sulpizio:GraphenePoisuille,Gallagher:grapheneCritical,Berdyugin:GrapheneCounterflow,Crossno:GrapheneWiedemannFranz,KuKim:grapheneFlowImaging}
(see Refs.~\cite{Narozhny:grapheneReview,Lucas:grapheneReview}
for a comprehensive review).

\begin{figure}[hb!]
	\includegraphics[width=0.7\linewidth]{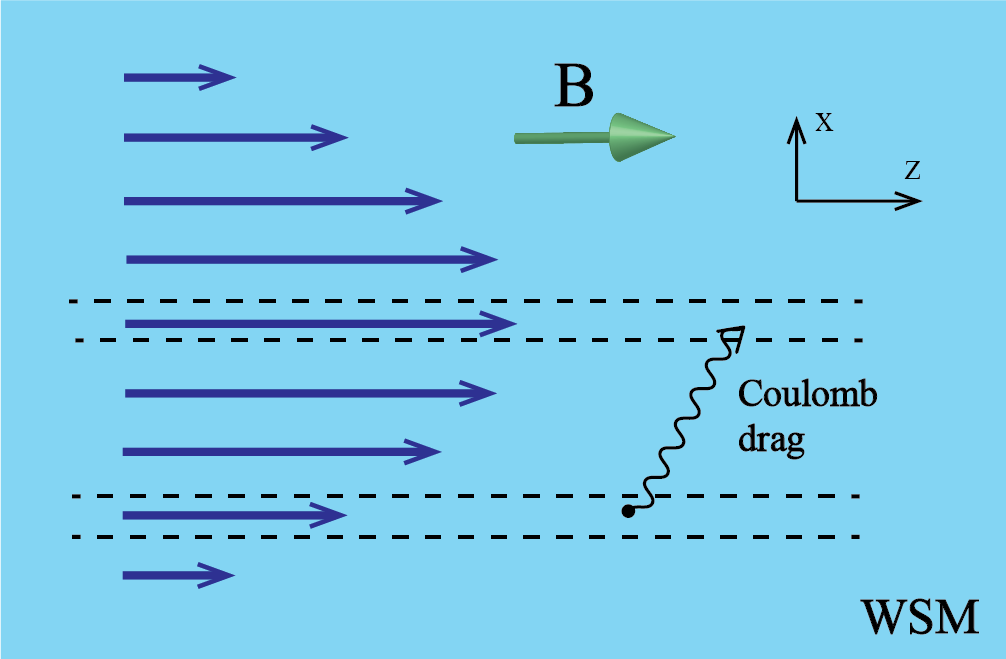}
	\caption{
		\label{fig:flowpicture}
		The flow of a Weyl liquid in a junction in a strong magnetic field.
		The motion is effectively unidirectional because quasiparticles can move only parallel or antiparallel to the magnetic field.
		The velocity of the liquid is defined as the velocity of the reference frame (bath) in which thermalisation of the liquid takes place.
		The dependence of the velocity on the transverse coordinate creates shear stress. The corresponding dissipative forces are determined by 
		the Coulomb interaction between different layers of the liquid.
	}
\end{figure}

In this paper, we demonstrate that 3D Weyl semimetals (WSMs) is a readily accessible platform for hydrodynamic transport and discuss manifestations of such
transport in them in strong magnetic fields. As demonstrated recently in Ref.~\cite{Bednik:WeylMagnetotransport}, the impurity scattering time $\tau$
for electrons in a Weyl semimetal with two nodes is strongly anisotropic as a function of the direction of the magnetic 
field:
\begin{align}
	\label{TimeAnisotropy}
	\frac{1}{\tau}=\frac{1}{\tau_0} \cos^2\theta+\frac{1}{\tau_1},
\end{align}
where $\theta$ is the angle between the field and the separation 
of the Weyl nodes in momentum space
and $1/\tau_0\gg 1/\tau_1$.
 The scattering rate is strongly suppressed for $\theta$ close to $\pi/2$, i.e. for magnetic fields perpendicular 
to the separation between the nodes, which makes the hydrodynamic regime in a Weyl semimetal 
conveniently achievable by applying the magnetic field in the respective direction (in addition to that, the elastic scattering
length in a strong magnetic field may be quite large even away from $\theta=\frac{\pi}{2}$; see Appendix~\ref{sec:EstimatesScatteringRate}).


In a strong magnetic field, Weyl electrons move predominantly (anti)parallel to the direction of the field.
As a result, the motion of the electron liquid in a Weyl-semimetal junction in a magnetic field resembles the Poiseuille flow~\cite{Suterra:PioseuilleReview,Landafshitz6} of
a liquid in a pipe,
as shown in Fig.~\ref{fig:flowpicture}.
The friction between the layers of the Weyl liquid moving with different velocities leads to dissipation and viscosity.
Due to the effectively one-dimensional character of the flow, there is no transverse transport of particles
 that leads to
a common mechanism of viscosity in liquids and gases~\cite{Reif:book}.
The viscosity comes, however, from the ``Coulomb drag'' between layers of liquid moving with different velocities, a mechanism
introduced recently for conventional metals in Ref.~\cite{LiaoGalitski:DragViscosity}.

We derive the hydrodynamic equations describing the hydrodynamic motion of a Weyl liquid in a strong magnetic field, compute microscopically the viscosity of
such a liquid and analyse the conduction of a Weyl-semimetal junction. We find that the viscosity weakly depends on the magnetic field and 
strongly on the temperature $T$ and, for realistic temperatures, is given by
\begin{align}
	\eta=\frac{M}{12\pi v_F}T^2,
	\label{ViscosityMain}
\end{align}
where the ``mass'' $M$ gives the inverse curvature of the quasiparticle dispersion near the Weyl nodes and $v_F$ is the Fermi velocity.
In the hydrodynamic regime, the thermal conductance (the response of the energy flux to a small temperature difference) of a Weyl-semimetal junction is
given, up to a non-universal coefficient of order unity which depends on the shape of the junction, by
\begin{align}
	G_q\sim \left(\frac{|e|B}{c}\right)^2\frac{S}{LMv_F}T,
	\label{ThermalConductance}
\end{align}
where $S$ is the cross-sectional area and $L$ is the length of the junction.

The paper is organised as follows.
In Sec.~\ref{sec:Model}, we introduce the model of WSMs in a strong magnetic field and discuss the approximations 
we use in this paper. In Sec.~\ref{sec:HydrodynamicEquations}, we derive the hydrodynamic equations for the electron liquid in such a semimetal.
Sec.~\ref{sec:viscosity} deals with the viscosity of such a liquid. In Sec.~\ref{sec:JunctionConductance}, we 
describe the hydrodynamic flow of the Weyl liquid generated by a temperature gradient and the possibility of its experimental observation.
We conclude in Sec.~\ref{sec:Conclusion}.


\section{Model}
\label{sec:Model}

\begin{figure}[ht!]
	\includegraphics[width=0.7\linewidth]{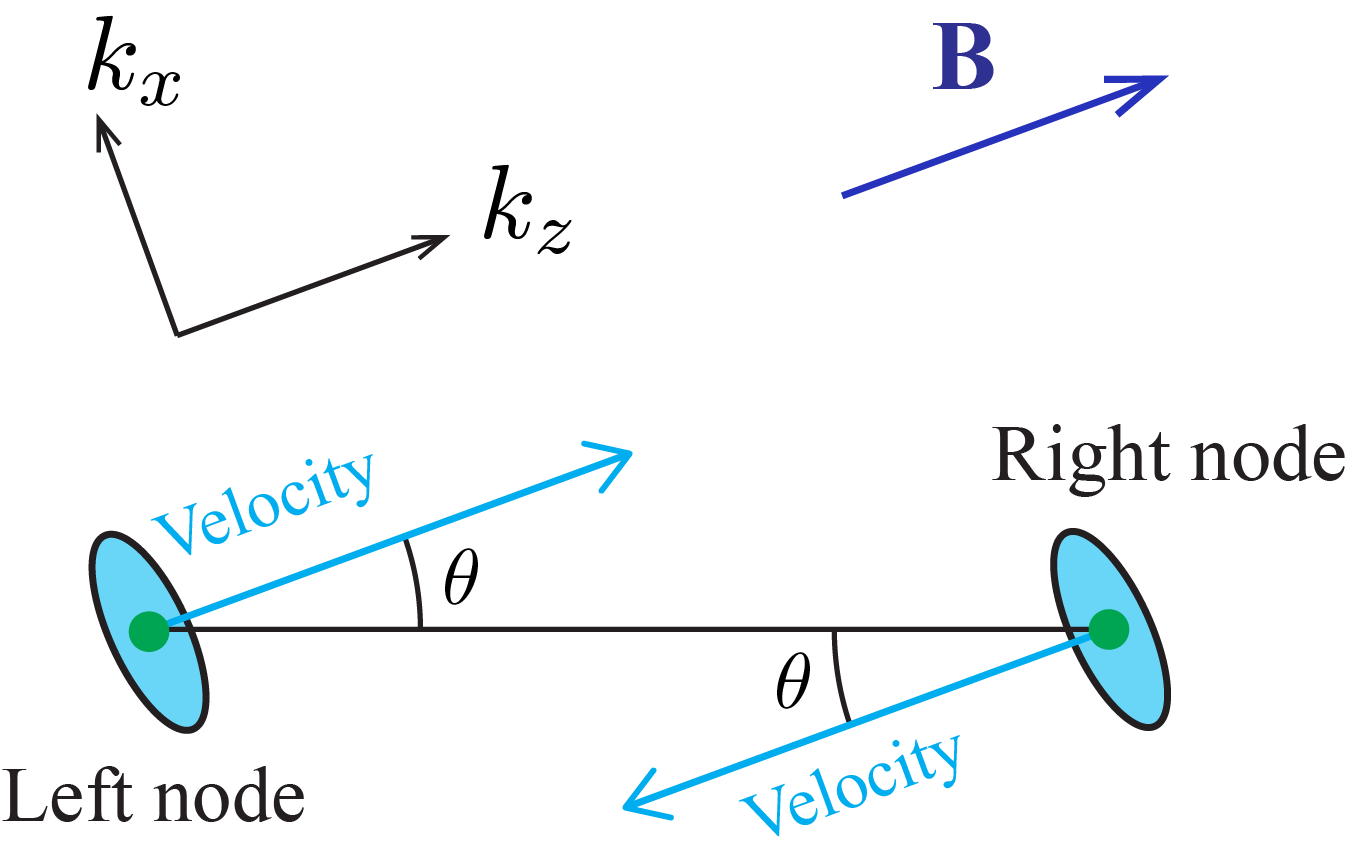}
	\caption{\label{fig:coordframes} Orientation of the magnetic field relative to the locations of the Weyl nodes in momentum space.}
\end{figure}

We consider the model of a Weyl semimetal with two Weyl nodes, right (R) and left (L) (shown in Fig.~\ref{fig:coordframes}),
and equal energies of the nodes. The Weyl semimetal may have additional nodes that weakly affect the electron dynamics between nodes L and R
so long as the additional nodes are well separated in momentum space from the nodes under consideration.
The magnetic field $\bB$ is directed at angle $\theta$ with respect to the line connecting the two nodes in momentum space.
For simplicity, we assume that the quasiparticles have no spin (apart from the pseudospin operator associated with the bands
in the Weyl semimetal) and have isotropic dispersions around each node. Our quantitative results will hold, however, up to coefficients
of order unity, for an arbitrary type-I Weyl semimetal.

We focus on the {\it ultraquantum limit} of the magnetic field
\begin{align}
	B>\left(\frac{2}{9}\right)^{\frac{1}{3}}\frac{\mu^2 c}{|e| v_F^2},
\end{align}
at which all electrons in equilibrium occupy the zeroth Landau level, where $\mu$ is the chemical potential (measured from the
energy of the Weyl nodes) in the absence of the field; $v_F$ is the Fermi velocity; hereinafter we set $\hbar=1$.

In the absence of impurity scattering and interactions, the motion of electrons is one-dimensional; quasiparticles can propagate
only parallel or antiparallel to the direction of the magnetic field $\bB$ with the velocity $v_F$. We assume that electrons move along the magnetic field
at node L and in the opposite direction at node R.

{\it Impurities and screening of Coulomb interactions.}
The strength of Coulomb interactions in the system is characterised by the dimensionless ``fine structure constant''
\begin{align}
	\alpha=\frac{e^2}{\epsilon v_F},
\end{align}
where $\epsilon$ is the dielectric constant. 
Most Weyl and Dirac materials have sufficiently large values of the dielectric constant~\cite{JAYGERIN1977771, Madelung, Jenkins:CarrierDynamics, Buckeridge:TaAsDielectricConstant} to ensure the condition $\alpha\ll 1$, which controls the diagrammatic perturbation theory for the 
interactions used in this paper.

 While we focus on the hydrodynamic regime of transport, 
the system may contain a small amount of charged impurities. The hydrodynamic transport, studied in this paper,
will persist so long as the elastic scattering rate~\eqref{TimeAnisotropy}
is significantly exceeded by the quasiparticle scattering rate due to electron-electron interactions.


In the Thomas-Fermi approximation~\cite{Abrikosov:metals}, the screening radius of static Coulomb interaction
is given by~\cite{Bednik:WeylMagnetotransport}
\begin{align}
	\varkappa^{-1}=\sqrt{\frac{\pi\epsilon v_F c}{2|e|^3 B}}=l_B\sqrt{\frac{\pi}{2\alpha}}.
	\label{ScreeningRadius}
\end{align}
The Thomas-Fermi approximation is justified in the limit~\cite{Abrikosov:metals} $|\mu|\gg \varkappa v_F$, which will be 
assumed throughout this paper. However, our results still hold qualitatively for other values of $\mu$.

{\it Spatial scales of the hydrodynamic flow.}
To apply the hydrodynamic description, we assume that the macroscopic degrees of freedom of the electron liquid, such as
the velocity $\bu$ of the liquid, vary smoothly 
in space, on length scales $\lambda$ significantly exceeding the microscopic scales of the system,
\begin{align}
	\label{SmoothApproximation}
	\lambda \gg Q^{-1}, l_B, \varkappa^{-1},
\end{align}
where $2Q$ is the separation between the Weyl nodes in momentum space;
the screening radius $\varkappa^{-1}$ is given by Eq.~\eqref{ScreeningRadius} and
\begin{align}
	l_B=\sqrt{c/(|e|B)}
	\label{MagneticLength}
\end{align}
is the magnetic length.
The momentum separation $2Q$ between the Weyl nodes is typically of the order of inverse atomic distances and is assumed to be
the largest momentum scale in the problem.

At low temperatures, the densities are correlated on length scales of the order of the screening radius~\eqref{ScreeningRadius}
$\varkappa^{-1}= l_B\sqrt{\pi/\alpha}$, 
which exceeds the magnetic length $l_B$ due to the smallness of the coupling constant $\alpha$.
The correlations in the electron liquid may, therefore, be assumed isotropic on lengthscales $\varkappa^{-1}\ll L\ll \lambda$ exceeding
the screening radius $\varkappa^{-1}$ but shorter than the characteristic scales $\lambda$ of the variation of the macroscopic hydrodynamic parameters
such as the velocity of the liquid.

{\it Energy scales.} 
As we demonstrate in this paper, the viscosity of the Weyl liquid strongly depends on its temperature.
For magnetic fields on the order of $1T$ or larger, the cyclotron frequency $v_F/l_B$ is on the order of $50$meV or larger and, thus, significantly exceeds the temperatures $T$ used in experiments on Weyl semimetals.
Taking into account Eq.~\eqref{ScreeningRadius}, realistic energy scales may, therefore, be assumed to satisfy the conditions\\
\begin{align}
	T\ll v_F \varkappa \ll v_F/l_B.
	\label{TemperatureSmallerEnergy}
\end{align}


\section{Hydrodynamic equations}

\label{sec:HydrodynamicEquations}

\subsection{Velocity of the hydrodynamic flow}
\label{sec:velocity}

In the hydrodynamic description, the flowing electron liquid may be considered to be equilibrated in a moving reference frame.
It is possible, therefore, to introduce the velocity $\bu$ of the electron liquid as the velocity of the equilibrium reference frame.
Considering the dispersion $\xi_\bk=\pm v_F k-u\left(\br\right)k$ of the quasiparticles at the zeroth Landau level near the left and the right nodes,
the respective quasiparticle occupation numbers are given by $f_L\left(k,u,\mu_L\right)=f\left(k,u,\mu_L\right)$ and $f_R\left(k,u,\mu_R\right)=f\left(-k,-u,\mu_R\right)$, where
\begin{align}
	\label{FDflowing}
	f\left(k,u,\mu\right)=\frac{1}{e^{\left[v_F k-u\left(\br\right)k-\mu\left({\bf r}\right)\right]/T}+1}
\end{align}
and the momentum $k$ is measured from the respective node. 
Hereinafter, we assume that the gradient of the velocity $\bu$ is small and neglect the effects of 
vorticity exemplified by the chiral vortical effect~\cite{Vilenkin:firstChiralVorticalEffect,Gorbar:WeylHydro,Chen:ChiralVorticalEffect}.

The electron liquid may be thermalised by any bath of neutral excitations (e.g. phonons) or the electrons themselves.
The full hydrodynamic description of a Weyl semimetal should include the hydrodynamic equations of motion of the bath as well
as those of the electron liquid. In this paper, we assume, for simplicity, that the electron liquid acts as its own bath.


\subsection{Hydrodynamic variables}

We develop a hydrodynamic description of the electron liquid in terms of the density of electrons near each node and the momentum density of the liquid.
The electron density $N_L$ near node L is measured relative to the equilibrium states of an undoped Weyl semimetals in the absence of the flow (i.e. for
$u=0$):
\begin{align}
\label{NL}
    N_L = \frac{|e|B}{2\pi c}\int\frac{dk}{2\pi}\left[f_L\left(k,u,\mu_L\right)-f_L\left(k,0,0\right)\right].
\end{align}
Using the distribution functions  $f_L\left(k,u,\mu_L\right)=f\left(k,u,\mu_L\right)$ given by Eq.~\eqref{FDflowing} we obtain
\begin{align}
	\label{NL1}
    N_L= \frac{|e|B }{4\pi^2 c}\frac{\mu_L}{v_F-u},
\end{align}
where $\mu_L$ is the chemical potential at the left node.

Because each quasiparticle at the left node moves with the velocity $v_F$ along the magnetic field
and carries a charge of $e=-|e|$, the electric current $j_L$ carried by the electrons near this node (relative to the equilibrium state
in the absence of the flow) is given by
\begin{align}
    j_L&= e\frac{|e| B}{2\pi c}\int\frac{dk}{2\pi}v_F\left[f_L\left(k,u,\mu_L\right)-f_L\left(k,0,0\right)\right]
    \nonumber\\
    &= ev_F N_L.
\end{align}

Similarly, we compute the concentration $N_R$ of the electrons near the right node:
\begin{align}
\label{NR}
	N_R=\frac{|e|B }{4\pi^2 c}\frac{\mu_R}{v_F+u}
\end{align}
and the current
\begin{align}
    j_R =-ev_F N_R.
\end{align}


\subsection{Hydrodynamic equations}

\subsubsection{Continuity equations for densities}

The continuity equation for the electron density $N_L$ reads
\begin{align}
\label{CNL}
	\partial_t N_L+v_F\partial_z N_L=
	-\frac{e^2}{4\pi^2 c}\bE\cdot\bB
    -\frac{N_L-N_R}{\tau},
\end{align}
where $v_F N_L$ is the flux of the density (measured relative to the equilibrium state) along the magnetic field ($z$ axis);
the first term in the right-hand side (rhs) describes the change of the density $N_L$ due to the chiral anomaly\cite{Burkov:review,SonSpivak:NegativeMR,Parameswaran:probingAnomaly,Burkov:AnomalyDiffusive} in the presence of the electric field $\bE$;
the second term in the rhs accounts for the elastic scattering of electrons between the two nodes. 
In Eq.~\eqref{CNL}, $1/\tau$ is the rate of internodal elastic scattering (due to collisions
with impurities or other defects in the system).
Similarly, the continuity equation for the density $N_R$ is given by
\begin{align}
\label{CNR}
	\partial_t N_R -v_F\partial_z N_R=
    \frac{e^2}{4\pi^2 c}\bE\cdot\bB-\frac{N_R-N_L}{\tau}.
\end{align}


\subsubsection{Navier-Stokes equation}

In order to provide a complete hydrodynamic description of the electron liquid in given electric and magnetic fields,
the continuity equations \eqref{CNL} and \eqref{CNR} have to be complemented by the Navier-Stokes equation
for momentum density. The momentum density near each individual Weyl node is not conserved due to the interactions between electrons
at different nodes.

The Navier-Stokes equation is given by
\begin{align}
	\label{CNV}
	\partial_{t}p+\partial_z J_p= F_{\text{E}}+F_{\text{scatt}}+F_{\text{visc}}-\partial_z P,
\end{align}
where $p$ is the density of momentum along the $z$ axis; $J_p$ is the flux of momentum; the force $F_{\text{E}}$ account for the change of the momentum $p$ due to external electric and magnetic fields; $F_{\text{scatt}}$ describes momentum relaxation due to impurity scattering; $F_{\text{visc}}$ is the force that describes dissipative effects
due to the viscosity of the electron liquid and $P$ is the pressure of the electron liquid. 
In what immediately follows, we compute these quantities microscopically in a weakly interacting Weyl electron liquid.

The momentum density is given by 
\begin{widetext}
	\begin{align}
		p =& \frac{|e|B}{2\pi c}\int\frac{dk}{2\pi} \left[f_L\left(k,u,\mu_L\right)-f_L\left(k,0,0\right)\right] \left(-Q\cos\theta+k\right)
		+ \frac{|e|B}{2\pi c}\int\frac{dk}{2\pi} \left[f_R\left(k,u,\mu_R\right)-f_R\left(k,0,0\right)\right] \left(Q\cos\theta+k\right)
		\nonumber\\
		= &N_L\left[-Q\cos\theta+\frac{\mu_L}{2\left(v_F-u\right)}\right]
		+N_R\left[Q\cos\theta-\frac{\mu_R}{2\left(v_F+u\right)}\right]
		+\frac{|e|BT^2}{6c}\frac{v_Fu}{(v_F^2-u^2)^2},
	\end{align}
where $2Q$ is the separation between the Weyl nodes in momentum space. The flux $J_p$ of momentum reads
\begin{align}
	\label{jp}
	J_{p}=&
	v_F\frac{|e|B}{2\pi c}\int\frac{dk}{2\pi} \left[f_L\left(k,u,\mu_L\right)-f_L\left(k,0,0\right)\right] \left(-Q\cos\theta+k\right)
	- v_F \frac{|e|B}{2\pi c}\int\frac{dk}{2\pi} \left[f_R\left(k,u,\mu_R\right)-f_R\left(k,0,0\right)\right] \left(Q\cos\theta+k\right)
	\nonumber\\
	=& v_F N_L \left[-Q\cos\theta+\frac{\mu_L}{2(v_F-u)}\right]
	-v_F N_R \left[Q\cos\theta-\frac{\mu_R}{2(v_F+u)}\right]
	+\frac{|e|BT^2}{12c}\frac{3v_F^2u^2-u^4}{\left(v_F^2-u^2\right)^2 v_F}.
\end{align}
Using Eq.~\eqref{NL1} and \eqref{NR}, the divergence of the flux $J_p$ can be simplified as
	\begin{align}
		\label{sjp}
		\partial_z J_p= & \left\{v_F \partial_z N_L \left[-Q\cos\theta+\frac{\mu_L}{2\left(v_F-u\right)}\right]
		-v_F\partial_z N_R\left[Q\cos\theta-\frac{\mu_R}{2\left(v_F+u\right)}\right]\right\}
		\nonumber\\
		&+\left[\frac{v_F N_L}{2}\partial_z \left(\frac{\mu_L}{v_F-u}\right)
		+\frac{v_F N_R}{2}\partial_z \left(\frac{\mu_R}{v_F+u}\right)\right]
		+v_F\frac{|e|B}{24c}\left[\partial_z \frac{T^2}{(v_F-u)^2}+\partial_z \frac{T^2}{(v_F+u)^2}\right]
		\nonumber\\
		= & v_F \partial_z N_L \left(-Q\cos\theta+\frac{\mu_L}{v_F-u}\right)
		-v_F\partial_z N_R\left(Q\cos\theta-\frac{\mu_R}{v_F+u}\right)
		\nonumber\\
		&+\frac{|e|BT^2}{6c}\frac{(3v_F^2u+u^3)v_F}{\left(v_F^2-u^2\right)^3}\partial_z u
		+\frac{|e|BT}{6c}\frac{3v_F^2u^2-u^4}{\left(v_F^2-u^2\right)^2 v_F}\partial_z T.
	\end{align}

The force $F_E$ is given by the change of the total momentum $p$ due to the transfer of quasiparticles between the nodes because of the chiral anomaly:
	\begin{align}
		F_{E}=& \frac{|e|B}{2\pi c}\partial_t\int \frac{dk}{2\pi} \left\{f_L\left[k,u,\mu_L-|e|E_z t\left(v_F-u\right)\right]-f_L\left(k,u,\mu_L\right)\right\}\left(-Q\cos\theta+k\right)\bigg|_{t=0}
		\nonumber\\
		&+\frac{|e|B}{2\pi c}\partial_t\int \frac{dk}{2\pi} \left\{f_R\left[k,u,\mu_R+|e|E_z t\left(v_F+u\right)\right]-f_R\left(k,u,\mu_R\right)\right\}\left(Q\cos\theta+k\right)\bigg|_{t=0}
	\end{align}
Using the distribution functions  $f_L\left(k,u,\mu_L\right)=f\left(k,u,\mu_L\right)$ given by Eq.~\eqref{FDflowing} we obtain
	\begin{align}
		\label{FE}
		F_E= &-\frac{e^2}{4\pi^2 c}\bE\cdot\bB\left(-Q\cos\theta+\frac{\mu_L}{v_F-u}\right)
		+\frac{e^2}{4\pi^2 c}\bE\cdot\bB\left(Q\cos\theta-\frac{\mu_R}{v_F+u}\right).
	\end{align}
In the limit of low temperatures $T$, Eq.~\eqref{FE} can be understood intuitively as follows.
The quantities $-Q\cos\theta+\frac{\mu_L}{v_F-u}$ and $Q\cos\theta-\frac{\mu_R}{v_F+u}$ give the momenta of the quasiparticles 
near the chemical potentials at the left and the right nodes and $\frac{e^2}{4\pi^2 c}\bE\cdot\bB$ is the rate of increase of
the quasiparticle density at the right
node (or its decrease at the left node). Multiplying these momenta by the corresponding rates of change of quasiparticle densities gives the rate 
of change of the total momentum due to an external electromagnetic field in the limit of zero temperature.
We emphasise, however, that the result \eqref{FE} applies at all temperatures $T$.

The momentum relaxation rate due to impurity scattering is given by
	\begin{align}
		\label{Fscatt}
		F_{scatt}=& \frac{|e|B}{2\pi c}\frac{1}{\tau}\int\frac{dk}{2\pi} \left[f_R\left(-k,u,\mu_R\right)-f_L\left(k,u,\mu_L\right)\right] \left(-Q\cos\theta+k\right)
		\nonumber\\
		&+ \frac{|e|B}{2\pi c}\frac{1}{\tau}\int \frac{dk}{2\pi} \left[f_L\left(-k,u,\mu_L\right)-f_R\left(k,u,\mu_R\right)\right] \left(Q\cos\theta-k\right)
		\nonumber\\
		= &\frac{N_L-N_R}{\tau}\left(2Q\cos\theta-\frac{\mu_L}{v_F-u}-\frac{\mu_R}{v_F+u}\right)
		-\frac{1}{\tau}\frac{|e|BT^2}{3c}\frac{uv_F}{\left(v_F^2-u^2\right)^2},
	\end{align}
\end{widetext}
where $1/\tau$ is the elastic internodal scattering rate introduced in Eqs.~\eqref{CNL} and \eqref{CNR}.
At $T=0$, Eq.~\eqref{Fscatt} can be understood intuitively as follows. 
At $T=0$, all the electron states with energies up to $\mu_L$ and $\mu_R$ are filled at the left and right nodes,
and it is possible to assume that only electrons with energies $\min(\mu_L,\mu_R)<\varepsilon<\max(\mu_L,\mu_R)$ get scattered between the nodes.
Then the quantities $-Q\cos\theta+\frac{1}{2}\left(\frac{\mu_L}{v_F-u}+\frac{\mu_R}{v_F+u}\right)$ and $Q\cos\theta-\frac{1}{2}\left(\frac{\mu_L}{v_F-u}+\frac{\mu_R}{v_F+u}\right)$ have the meaning of the average momenta of electrons
at the left and the right nodes that participate in these elastic scattering processes.
Multiplying these momenta by the rate $\frac{N_L-N_R}{\tau}$ of change of the densities of electrons due to internodal scattering
gives Eq.~\eqref{Fscatt} at $T=0$.

The force $F_{visc}=\partial_x T_{xz}+\partial_y T_{yz}+\partial_z T_{zz}$ describes the dissipative effects due to the viscosity
of the liquid, where $T_{xz}$, $T_{yz}$ and $T_{zz}$ are the components of the stress tensor.
While the liquid can move only along the direction of the magnetic field (the $z$ axis, see Fig.~\ref{fig:flowpicture}) in the strong magnetic field under consideration,
the velocity $u$ of this motion is different for different transverse coordinates $x$ and $y$ for the same $z$, which creates shear stress.
The total viscous force
is given by
\begin{align}
	\label{Fvisc}
	F_{visc}=\eta\left(\partial_x^2+ \partial_y^2\right)u+\kappa \partial_z^2 u,
\end{align}
where $u$ is the velocity of the electron liquid defined in Sec.~\eqref{sec:velocity};
$\eta$ is the shear viscosity, the response of the stress forces between layers of the electron liquid flowing along the $z$ axis
to the transverse gradient of the velocity $u$; the coefficient $\kappa$
characterises the response of the strain to the longitudinal spatial change
of the velocity $u$. We microscopically demonstrate in Sec.~\eqref{sec:viscosity}
that $\eta\gg \kappa$.

The pressure $P$ of the Weyl liquid, computed in Appendix~\ref{sec:Pressure}, is given by
\begin{align}
	\label{PressureMain}
	P=P_0+\frac{|e|B}{12 cv_F}T^2,
\end{align}
where $P_0$ is a temperature-independent contribution that depends on the details of the quasiparticle 
dispersion away from the Weyl nodes.

\begin{widetext}

Combining Eqs.~\eqref{CNV}, \eqref{sjp}, \eqref{FE}, \eqref{Fscatt} and \eqref{Fvisc},
we arrive at the
Navier-Stokes equation ~\eqref{CNV} in the form

\begin{align}
	\label{CNV1}
	&\partial_t p + v_F \partial_z N_L \left(-Q\cos\theta+\frac{\mu_L}{v_F-u}\right)
	-v_F\partial_z N_R\left(Q\cos\theta-\frac{\mu_R}{v_F+u}\right)
	\nonumber\\
		&+\frac{|e|BT^2}{6c}\frac{(3v_F^2u+u^3)v_F}{\left(v_F^2-u^2\right)^3}\partial_z u
		+\frac{|e|BT}{6c}\frac{3v_F^2u^2-u^4}{\left(v_F^2-u^2\right)^2 v_F}\partial_z T =
	\nonumber\\
	&-\frac{e^2}{4\pi^2 c}\bE\cdot\bB\left(-Q\cos\theta+\frac{\mu_L}{v_F-u}\right)
	+\frac{e^2}{4\pi^2 c}\bE\cdot\bB\left(Q\cos\theta-\frac{\mu_R}{v_F+u}\right)
	\nonumber\\
	&+\frac{N_L-N_R}{\tau}\left(2Q\cos\theta-\frac{\mu_L}{v_F-u}-\frac{\mu_R}{v_F+u}\right)
	-\frac{1}{\tau}\frac{|e|BT^2}{3c}\frac{uv_F}{\left(v_F^2-u^2\right)^2}
	+\eta\left(\partial_x^2+ \partial_y^2\right)u+\kappa \partial_z^2 u-\partial_z P.
\end{align}
Equations~\eqref{CNL}, \eqref{CNR} and \eqref{CNV1} provide a complete hydrodynamic description of a Weyl liquid in a strong magnetic field.

\end{widetext}


\section{Viscosity}

\label{sec:viscosity}

In this section, we compute microscopically the viscosity of a Weyl liquid in a strong magnetic field.
The viscosity tensor is determined by the correlator of the corresponding components 
of the stress tensor (see, for example, Ref.~\cite{Bradlyn:KuboFormula}) and can be represented in the form
\begin{align}
	\eta_{ijkl}(\omega)=\frac{1}{\Omega}\left[\cB_{ijkl}(\omega)-\cB_{ijkl}(0)\right],
	\label{ViscosityTensor}
\end{align}
where 
\begin{align}
	&\cB_{ijkl}(\omega)= \nonumber\\
	&\frac{1}{2}\int d\br\int_{-\beta}^{\beta}d\tau \left<T_\tau\hT_{ij}(\br,\tau)\hT_{kl}(\bzero,0)\right>e^{i\Omega\tau} \bigg|_{i\Omega\rightarrow \omega+i0}
	\label{Correlator}
\end{align}
is the retarded correlator of the components $\hT_{ij}$ and $\hT_{kl}$ of the stress tensor operator and
$i\Omega\rightarrow \omega+i0$ is our convention for the analytic continuation from positive Matsubara frequencies $\Omega$ to the real frequency $\omega$~\cite{Mahan:book,AGD}.

Strictly speaking, the viscosity of the electron liquid depends on the velocity $\bu$ of the liquid at a given location, and the averaging 
$\left<\ldots\right>$ in Eq.~\eqref{Correlator} should be carried out with respect to the equilibrium Fermi-Dirac distribution \eqref{FDflowing}
in the reference frame of the moving liquid. However, because realistic velocities $\bu$ are significantly exceeded by the Fermi velocity $v_F$,
the dependence of the viscosity on the velocity $\bu$ may be neglected and averaging over the equilibrium state of a stationary liquid may be used when 
computing the viscosity tensor~\eqref{ViscosityTensor}.
In what follows, we evaluate explicitly the Matsubara correlator in Eq.~\eqref{Correlator}.

The stress tensor $\hT_{ij}$ includes two qualitatively distinct components~\cite{Schwinger:ManyParticleSystems}.
The first, kinetic, component is independent of the interaction in the system and for a Weyl semimetal with two nodes, is given by
\begin{align}
	\label{KineticStressTensor}
	\hT_{ij}^{(0)}(\br)=\sum_{\chi=L,R}\hpsi^\dagger_\chi(\br) {\hat v}_{i\chi}\hk_j \hpsi_\chi(\br),
\end{align}
where the summation is carried out over the nodes $\chi=L,R$;
$\hpsi^\dagger_\chi(\br)$ and $\hpsi_\chi(\br)$ are the creation and annihilation operators of the electrons at node $\chi$;
${\hat v}_{i\chi}$ is the $i$-th component of the velocity operator at node $\chi$
and $\hk_j=-i\frac{\partial}{\partial r_j}$ is the $j$-th momentum component.
The second contribution to the stress tensor $\hT_{ij}$ is determined by the 
electron-electron interactions~\cite{Schwinger:ManyParticleSystems} (see also Refs.~\cite{LinkSchmalian:grapheneStressTensor}
and \cite{LiaoGalitski:DragViscosity}) and, in the limit of smooth variations of the gradients of the 
macroscopic parameters of the liquid [cf. the condition~\eqref{SmoothApproximation}] (``local uniformity approximation'' of Ref.~\cite{Schwinger:ManyParticleSystems}), is given by
\begin{align}
	\label{InteractionStressTensor}
	\hT_{ij}^{\text{(int)}} (\br) = - \frac{1}{2}\sum_{\chi,\chi^\prime=L,R} \int d \brho\,\,
	\hpsi^{\dagger}_\chi \left(\br + \frac{\brho}{2}\right)  \hpsi^{\dagger}_{\chi^\prime} \left(\br - \frac{\brho}{2}\right)
	\nonumber\\
	\rho_i \frac{\partial V(\brho)}{\partial \rho_j}
	\hpsi_{\chi^\prime} \left(\br - \frac{\brho}{2}\right)  \hpsi_\chi \left(\br + \frac{\brho}{2}\right),
\end{align}
where $V(\brho)=\frac{e^2}{\varepsilon \rho}$ is the Coulomb interaction potential.


\subsection{Shear viscosity}

In what immediately follows, we compute the
viscosity $\eta=\eta_{xzxz}=\eta_{yzyz}$ that describes the response of the shear stress~\cite{Landafshitz7}
$T_{xz}$ and $T_{yz}$ of the liquid
flowing along the $z$ axis, the direction of the magnetic field (see Fig.~\ref{fig:flowpicture}), to the transverse gradients $\frac{\partial u}{\partial x}$
and $\frac{\partial u}{\partial y}$ of the velocity.
Because the quasiparticles at both nodes can move only (anti-)parallel to the magnetic field ($v_{x\chi}=v_{y\chi}=0$, $v_{z\chi}\neq0$),
there is no kinetic contribution~\eqref{KineticStressTensor} to the components $T_{xz}$ and $T_{yz}$ of the stress tensor, which determine the 
viscosity $\eta$.
The absence of such a contribution is due to the effectively one-dimensional character of the motion of the liquid.
Indeed, in conventional liquids and gases, in which interactions between particles may be considered to be contact,
the viscosity comes from the transverse transport of particles between parallel moving layers of liquid and the transfer of the longitudinal momentum of those
 particles~\cite{Reif:book}. Because transverse transport is negligible in a Weyl semimetal in a strong magnetic 
field, the viscosity is dominated by the Coulomb drag between parallel moving layers of liquid, which comes from the long-range 
character of (screened) Coulomb interactions.
In what follows, we compute, therefore, the Matsubara correlator [cf. Eq.~\eqref{Correlator}] of the interaction contributions~\eqref{InteractionStressTensor} to the stress tensor. 

The electron liquids may relax momentum via processes of quasiparticle scattering between the nodes.
Due to the long-range nature of Coulomb interactions, with the characteristic momentum scale $\varkappa$ given
by Eq.~\eqref{ScreeningRadius}, such processes have a rate suppressed by the small parameter $\varkappa/ Q \ll 1$ and will not 
be considered here.

Another possible mechanism of viscosity comes from the Coulomb drag~\cite{Pogrebinskii:FirstDrag,Price:FirstDrag,NarozhnyLevchenko:CoulombDrag}
between layers of the electron liquid moving parallel to each other, as shown in Fig.~\ref{fig:flowpicture}.
In the presence of the transverse gradient of the velocity $\bu$, different layers of the electron liquid move with different velocities, with Coulomb 
interactions resulting in effective friction forces between the layers.
This mechanism of viscosity has been pioneered in Ref.~\cite{LiaoGalitski:DragViscosity} for 
a conventional Fermi liquid. Under the made approximations, it also dominates the viscosity of Weyl fermions in a strong magnetic
field considered here.

\begin{figure}[H]
	\centering
	\includegraphics[width=1\linewidth]{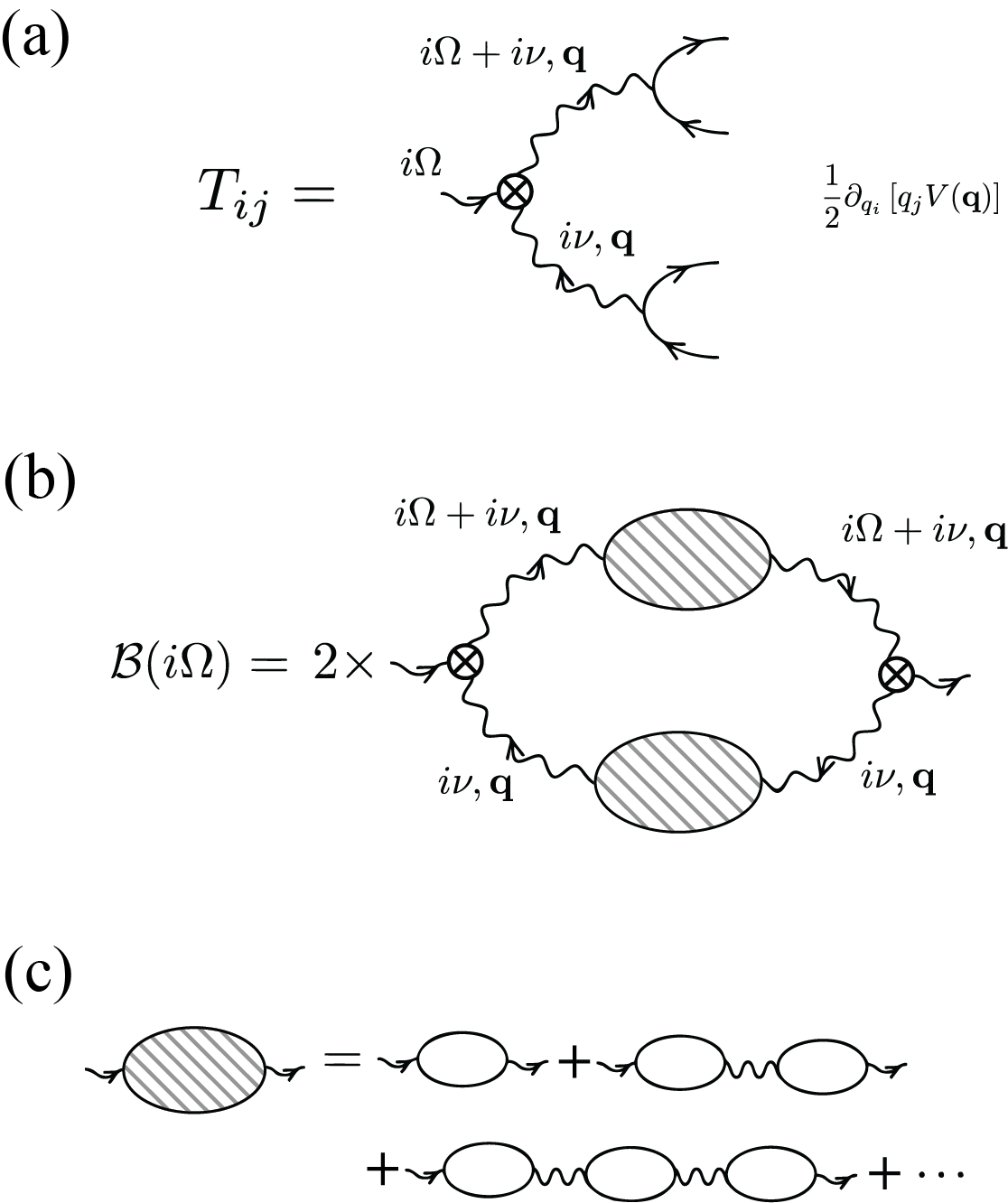}
	\caption{\label{fig:viscositydiagrams} 
	Diagrams for computing the viscosity of the electron liquid. 
	(a) The vertex corresponding to the interaction contribution to the stress tensor.
	(b) The diagram for the drag contribution to the viscosity. 
	(c) A block in diagram (b) which takes into account the screening of the interactions in the RPA.}
\end{figure}

The drag contribution to the Matsubara correlator $\cB_{ijkl}(i\Omega)$ in Eq.~\ref{Correlator} corresponds
to the diagram in Fig.~\ref{fig:viscositydiagrams}b, where the interaction contribution
to the stress tensor $T_{ij}$ corresponds to the vertex shown in Fig.~\ref{fig:viscositydiagrams}a. To describe the screening of the interactions,
we use the random phase approximation (RPA)~\cite{Mahan:book,Abrikosov:metals}, as shown in Fig.~\ref{fig:viscositydiagrams}c.
A prefactor of $2$ in diagram~\ref{fig:viscositydiagrams}b comes from two possible pairings of the ends of the stress-tensor vertex~\ref{fig:viscositydiagrams}a
in the correlator that this diagram describes.

The correlator $\cB_{ijkl}(i\Omega)$, corresponding to the diagram~\ref{fig:viscositydiagrams}b,
can be evaluated in the momentum representation as
\begin{align}
	\cB_{ijkl}(i\Omega)=
	&\frac{T}{2}\sum_{i\nu}\int \frac{d^3\bq}{(2\pi)^3}
	\frac{\partial}{\partial q_i}\left[q_j V(\bq)\right]
	\frac{\partial}{\partial q_k}\left[q_l V(\bq)\right]
	\nonumber\\
	&\frac{\Pi\left(i\Omega+i\nu,\bq\right)}{1-V(\bq)\Pi\left(i\Omega+i\nu,\bq\right)}
	\frac{\Pi\left(i\nu,\bq\right)}{1-V(\bq)\Pi\left(i\nu,\bq\right)},
	\label{CorrelatorMatsubara}
\end{align}
where 
$V(\bq)=\frac{4\pi e^2}{\epsilon q^2}$ is the bare propagator of Coulomb interactions and
$\Pi(i\nu, q)$ is the polarisation operator, which corresponds
to a simple fermionic 
bubble in the diagrams in Fig.~\ref{fig:viscositydiagrams}c in the limit of a small coupling $\alpha\ll1$.
A microscopic calculation of the polarisation operator are presented in Appendix~\ref{sec:PolarisationOperator}.

Equation~\eqref{CorrelatorMatsubara} contains a Matsubara sum of the form $I(i\Omega)=T\sum_{i\nu} D(i\nu+i\Omega)D(i\nu)$,
which can be conveniently computed by contour integration in the complex $\nu$ plane that gives
\begin{widetext}
	\begin{align}
		I(i\Omega)=&\frac{1}{4\pi i}\int_{-\infty}^{+\infty} \coth\frac{\varepsilon}{2T} \left[D_A(\varepsilon+i\Omega)D_R(\varepsilon)
		-D_R(\varepsilon+i\Omega)D_A(\varepsilon)\right]d\varepsilon
		\nonumber\\
		&+\frac{1}{4\pi i}\int_{-\infty}^{+\infty} \coth\frac{\varepsilon}{2T} \left[D_R(\varepsilon)D_R(\varepsilon-i\Omega)
		-D_A(\varepsilon)D_A(\varepsilon-i\Omega)\right]d\varepsilon,
		\nonumber
	\end{align}
\end{widetext}
where $D_A(\varepsilon)$ and $D_R(\varepsilon)\equiv D_A^*(\varepsilon)$ are the advanced and retarded versions of the correlator $D(i\nu)$, i.e.
obtained from it by analytic continuation from, respectively, the lower and the upper half-planes (see, for example, Refs.~\cite{PerelEliashberg:analyticContinuation}
and \cite{KamenevOreg:drag} for the details of the contour integration).
Performing such contour integration and the analytic continuation $i\Omega\rightarrow \omega+i0$ and utilising Eq.~\eqref{ViscosityTensor} gives, in the limit
of low frequencies $\omega$,
\begin{align}
	&\eta(\omega\rightarrow0)=
	\nonumber\\
	&\frac{1}{T}\int \frac{d\varepsilon}{2\pi}\frac{d^3\bq}{(2\pi)^3}
	\left[\frac{q_x q_z V^\prime(q)}{2q\sinh\frac{\varepsilon}{2T}}\right]^2
	\nonumber\\
	&\frac{ \left[\mathrm{Im}\, \Pi_R(\varepsilon, \bq) \right]^2 }{\left\{\left[ 1 - V(\bq)\, \mathrm{Re}\, \Pi_R(\varepsilon, \bq)\right]^2
		+ \left[V(\bq)\, \mathrm{Im}\,
		\Pi_R(\varepsilon, \bq)\right]^2 \right\}^2},
	\label{ViscosityWorking}
\end{align}
where $\Pi_R(\varepsilon, \bq)$ is the retarded polarisation operator obtained by analytic continuation from the Matsubara polarisation operator 
$\Pi(i\nu, \bq)$ given by Eq.~\eqref{PolarisationOperatorDefinition}.
Equation~\eqref{ViscosityWorking} has been obtained in Ref.~\cite{LiaoGalitski:DragViscosity} using Keldysh technique.
In what immediately follows, we evaluate explicitly the polarisation operator for experimentally important frequency and momentum scales.


\subsubsection*{Polarisation operator}

The polarisation operator
\begin{align}
	\label{PolarisationOperatorDefinition}
	\Pi(\br;\br^\prime;i\Omega)=-\frac{1}{2}\int_{-\beta}^{\beta}\left<\hn(\br,\tau)\hn(\br^\prime,0)\right>e^{i\Omega\tau}d\tau,
\end{align}
where $\hn(\br,\tau)$ is the electron density,
is evaluated explicitly in  Appendix~\ref{sec:PolarisationOperator}.
In the limit $\varkappa^{-1}\gg l_B$ under consideration and at momenta $|\bq|\ll l_B^{-1}$,
the Fourier-transform of the retarded polarisation operator is given by
\begin{align}
	&\Pi_R(\varepsilon, \bq)=\nonumber\\
	&\frac{|e|B}{2 \pi c} 
	\sum_{\chi = L, R}
	\int \frac{dk}{2 \pi} \frac{f_\chi(k,0,\mu_\chi) -  f_\chi(k+q_z,0,\mu_\chi)}
	{E_{\chi k} - E_{\chi (k+q_z)}+\varepsilon+i0},
	\label{PolarisationOperatorWorking}
\end{align}
where the integration is carried out over the momentum $k$ along the direction of the magnetic field;
$f_\chi(k,0,\mu_\chi)$ is the distribution function of the electrons at node $\chi$ [cf. Eq.~\eqref{FDflowing}] and
$E_{\chi k}$ is the corresponding electron dispersion (as we clarify below, the deviation of the dispersion from 
the linear dependence $\pm v_F k$ needs to be taken into account for evaluating the viscosity of the system).

The real and imaginary parts of the retarded polarisation operator $\Pi_R(\varepsilon, \bq)$
describe, respectively, the screening of Coulomb interactions and the decay of the density waves in the 
Weyl liquid (Landau damping). In what immediately follows, we evaluate these contributions explicitly.

\subsubsection*{Screening}

The main contribution to the static viscosity~\eqref{ViscosityWorking} comes from the energies
$\varepsilon$ on the order of the temperature $T$, which is significantly exceeded by the cyclotron frequency [see Eq.~\eqref{TemperatureSmallerEnergy}]
and the bandwidth of the quasiparticle dispersion.
This allow us to neglect the
$\varepsilon$-dependence of the real part $\text{Re} \Pi_R$ of the
retarded polarisation operator.

Similarly, we neglect the dependence of  $\text{Re} \Pi_R$ on the momentum $\bq$ whose characteristic values are on the order
of the inverse screening radius $\varkappa$ given by Eq.~\eqref{ScreeningRadius} and significantly exceeded by the
inverse magnetic length $l_B^{-1}$ [cf. the condition~\eqref{TemperatureSmallerEnergy}] and the momentum scales of the
quasiparticle band. Below, we will show that the typical scale of the momentum component $q_z$ that contributes to the viscosity
is even smaller and is on the order of $T/v_F$.

The real part of the polarisation operator is, therefore, given by the density of the electron states at the Fermi level (with the minus sign):
\begin{align}
	\text{Re}\, \Pi_R (\varepsilon, \bq)=-\frac{|e|B}{2\pi^2 cv_F}=-\frac{\epsilon \varkappa^2}{4\pi e^2},
	\label{RePi}
\end{align}
where $\varkappa$ is the inverse screening radius of Coulomb interactions given by Eq.~\eqref{ScreeningRadius}.

\subsubsection*{Landau damping}

For the existence of a finite damping (to the leading order in interactions), it is 
necessary to take into account the curvature of the electron dispersion near the nodes.
Indeed, for linearly dispersive quasiparticles, 
density waves composed of electrons near one node propagate with the velocity $\pm v_F$ and lack dispersion.
The conservation of momentum in any process involving only electrons near one node also enforces energy conservation, which is why
all momentum conserving processes contribute to the damping and lead to a singular $\propto\sum_\pm\delta(\varepsilon \pm v_F q_z)$ imaginary part of 
the lowest-order polarisation operator~\eqref{PolarisationOperatorWorking}.

In order to describe a finite dispersion of the charge density waves, we take into account the non-linearity of the quasiparticle dispersion near the Weyl nodes:
\begin{eqnarray}
	E_{k} = \pm v_F k + \frac{k^2}{2M},
	\label{DispersionNonLinear}
\end{eqnarray}
where ``$+$'' and ``$-$'' correspond, respectively, to the left and the right nodes.
The dispersion \eqref{DispersionNonLinear} and the momentum $k$ are measured, respectively, from the Fermi level 
and Fermi momentum.
The energy scale $Mv_F^2$ 
 is the largest
energy scale in the problems and, in the case it is determined by the band structure of the Weyl semimetal, may be assumed to 
be on the order of several electronvolt.

Utilising Eqs.~\eqref{PolarisationOperatorWorking} and \eqref{DispersionNonLinear} gives
\begin{align}
	&\text{Im}\, \Pi_R (\varepsilon, \bq)
	=\frac{|e|B}{4\pi c}\frac{M}{|q_z|}
	\sum_\pm
	\nonumber \\
	&\frac{\pm \sinh\frac{\varepsilon}{2T}}
	{\cosh\left[\frac{Mv_F^2(\varepsilon\pm v_F q_z)}{2\varepsilon T}+\frac{\varepsilon}{4T}\right]
		\cosh\left[\frac{Mv_F^2(\varepsilon\pm v_F q_z)}{2\varepsilon T}-\frac{\varepsilon}{4T}\right]
	}.
	\label{LandauDamping}
\end{align}
The terms with ``+'' and ``-'' correspond to functions sharply peaked at $\varepsilon\pm v_F q_z=0$ and account, respectively,
for the contribution of the left and right nodes.

The characteristic values $\varepsilon$ and $v_F q_z$ that contribute to the viscosity are on the order of the temperature $T$.
When deriving Eq.~\eqref{LandauDamping} we neglected, therefore, the effect of the small energy $\frac{q_z^2}{2M}\sim \frac{T^2}{Mv_F^2}\ll 1$ on the distribution functions 
$f_\chi(k,0,\mu_\chi)$ and  $f_\chi(k+q_z,0,\mu_\chi)$
 in the polarisation operator~\eqref{PolarisationOperatorWorking}.


\subsubsection*{The value of the shear viscosity}

Utilising Eqs.~\eqref{ViscosityWorking} and \eqref{RePi} and the smallness of the Landau damping, the viscosity can be rewritten in the form
\begin{align}
	\eta=\frac{1}{T}\left(\frac{4\pi e^2}{\epsilon}\right)^2
	\int\frac{d\varepsilon}{2\pi}\frac{d^3\bq}{(2\pi)^3}
	\frac{q_x^2 q_z^2}{\sinh^2\frac{\varepsilon}{2T}}
	\frac{\left[\mathrm{Im}\, \Pi_R(\varepsilon, \bq) \right]^2}{\left(q^2+\varkappa^2\right)^4}.
	\label{ViscosityIntermediate1}
\end{align}
Because the imaginary part $\text{Im}\, \Pi_R (\varepsilon, \bq)$ of the retarded polarisation operator 
is sharply peaked at $\varepsilon\pm vq_z=0$ only momenta $q_z$ on the order of $T/v_F$ contribute to the viscosity.
By contrast, the transverse momenta $q_x$ and $q_y$ have characteristic values on the order of $\varkappa$, which significantly 
exceed $T/v_F$ [see the condition \eqref{TemperatureSmallerEnergy}]. This allows us to neglect the dependence of the denominator in
Eq.~\eqref{ViscosityIntermediate1} on the momentum $q_z$. Integrating out the transverse momenta $q_x$ and $q_y$ gives
\begin{align}
	\eta=\frac{1}{3\pi T\varkappa^4}\left(\frac{\pi e^2}{\epsilon}\right)^2
	\int\frac{d\varepsilon}{2\pi}\frac{d q_z}{2\pi}
	\frac{\left[\mathrm{Im}\, \Pi_R(\varepsilon, \bq) \right]^2}{\sinh^2\frac{\varepsilon}{2T}}q_z^2.
	\label{ViscosityIntermediate2}
\end{align}

Using Eq.~\eqref{LandauDamping} and introducing variables $s=\frac{\varepsilon}{4T}$ and $t=\frac{Mv_F^2(v_F q_z-\varepsilon)}{2\varepsilon T}$,
Eq.~\eqref{ViscosityIntermediate2} can be represented in the form
\begin{align}
	\eta=&\frac{1}{3\pi^3 T\varkappa^4}\left(\frac{\pi e^2}{\epsilon}\right)^2\left(\frac{|e|B}{4\pi c}\right)^2
	\nonumber\\
	&\frac{16 T^3M}{v_F^3}\int\frac{|s| dt ds}{\cosh^2(s+t)\cosh^2(s-t)},
\end{align}
which gives the viscosity 
\begin{align}
	\eta=\frac{M}{12\pi v_F}T^2.
	\label{ViscosityFinal}
\end{align}
Equation~\eqref{ViscosityFinal} is our main result for the viscosity of a Weyl liquid in a strong magnetic field.

Due to the drag character of the analyzed mechanism of viscosity, it
has the same temperature dependence, $\propto T^2$, as the drag resistivity between two parallel conductive layers~\cite{NarozhnyLevchenko:CoulombDrag}
and can be understood from phase-space considerations similar to those explaining the $\propto T^2$ dependence of the electron-electron scattering rate
in a conventional metal~\cite{Abrikosov:metals,Gantmakher:book}.
Indeed, each electron can collide with electrons in a parallel layer of the liquid with energies in a window of order $T$ near the Fermi energy.
The characteristic width of the layer of momenta into which the electron can get scattered is also proportional to the temperature $T$.
In the presence of a finite curvature $M^{-1}$ of the dispersion, the final momentum of the other electron participating in the collision
is fixed by the energy and momentum conservation laws. This results, therefore, in the $\propto T^2$ scattering rate that manifests itself 
in the viscosity \eqref{ViscosityFinal}.


\subsection{Longitudinal response}

In this subsection, we evaluate the coefficient $\kappa$ that characterises 
the response of stress component $T_{zz}$ to the gradient $\frac{\partial u}{\partial z}$ [cf. Eq.~\eqref{Fvisc}].
This coefficient has both drag and kinetic contributions.

\begin{align}
	\kappa^{\text{drag}}=\frac{1}{T}\left(\frac{4\pi e^2}{\epsilon}\right)^2
	\int\frac{d\varepsilon}{2\pi}\frac{d^3\bq}{(2\pi)^3}
	\frac{q_z^4}{\sinh^2\frac{\varepsilon}{2T}}
	\frac{\left[\mathrm{Im}\, \Pi_R(\varepsilon, \bq) \right]^2}{\left(q^2+\varkappa^2\right)^4}.
	\label{ViscosityIntermediate3}
\end{align}

Using Eq.~\eqref{LandauDamping} and introducing variables $s=\frac{\varepsilon}{4T}$, $t=\frac{Mv_F^2(v_F q_z-\varepsilon)}{2\varepsilon T}$ and $b=\frac{4T}{v_F}$,
Eq.~\eqref{ViscosityIntermediate3} can be represented in the form
\begin{align}
	\kappa^{\text{drag}}=\frac{M}{12\pi v_F}T^2\cdot \frac{4}{\varkappa^2} \int\frac{\left(bs\right)^2|s|ds dt}{\cosh^2\left(s-t\right)\cosh^2\left(s+t\right)}.
	\label{ViscosityIntermediate4}
\end{align}
By introducing Eq.~\eqref{ViscosityFinal}, Eq.~\eqref{ViscosityIntermediate4} can be expressed as
\begin{align}
	\kappa^{\text{drag}}&=\frac{4}{\varkappa^2} \int\frac{\left(bs\right)^2|s|ds dt}{\cosh^2\left(s-t\right)\cosh^2\left(s+t\right)} \eta
	\nonumber \\
	&=\frac{3\zeta\left(3\right)b^2}{\varkappa^2}\eta  \sim \frac{M}{v_F^3 \varkappa^2}T^4.
	\label{ViscosityIntermediate5}
\end{align}

For the low temperatures $T\ll \varkappa v_F$ under consideration [cf. Eq.~\eqref{TemperatureSmallerEnergy}], the contribution \eqref{ViscosityIntermediate5} is significantly exceeded by the shear viscosity~\eqref{ViscosityFinal}.
The contribution \eqref{ViscosityIntermediate5} is proportional to the typical value of $q_z^4\propto T^4$ and, as a result, contains an extra power of $T^2$
relative to the shear viscosity \eqref{ViscosityFinal}. It is possible to show that the kinetic contribution [i.e. containing the kinetic vertices \eqref{KineticStressTensor}] has the same dependence on the typical value of $q_z$ and the $\propto T^4$ temperature dependence and is also
significantly suppressed compared to the shear viscosity $\eta$. We leave, however, a
rigorous calculation of this contribution for future studies.


\section{Temperature-generated flow and potential for experimental observation}

\label{sec:JunctionConductance}


In this section, we address the possibility of experimental observation of the discussed hydrodynamic flow of a Weyl electron liquid 
in a strong magnetic field. In a sufficiently long Weyl-semimetal junction, whose length exceeds the elastic scattering length
$\tau v_F$,
the conductance is independent of the viscosity $\eta$. Indeed, according to Eqs.~\eqref{CNL} and \eqref{CNR}, a longitudinal electric field $E$
results in a stationary imbalance of the electron densities $N_L-N_R=-\frac{e^2}{4\pi^2c}\bB\cdot\bE \tau$, which leads to a finite conductivity
$\sigma=\frac{|e|^3v_F}{4\pi^2 c}B\tau$ matching the conductivity in a system in the non-hydrodynamic (diffusive) regime~\cite{SonSpivak:NegativeMR,Burkov:review}.

The hydrodynamic properties of the systems, however, manifest themselves in heat transport.
The hydrodynamic flow can be generated by a temperature gradient and detected through the dependence of the heat flux 
on the temperature and magnetic field.

For a stationary flow, the momentum flux and electron densities at nodes $L$ and $R$ do not change, $\partial_t p=\partial_t N_L =\partial_{t} N_R=0$.
Multiplying the continuity equations~\eqref{CNL}
and \eqref{CNR} by, respectively, $-Q\cos\theta+\frac{\mu_L}{v_F-u}$ and $Q\cos\theta-\frac{\mu_R}{\left(v_F+u\right)}$ and subtracting from 
the Navier-Stokes equation \eqref{CNV1} gives
\begin{align}
	&\label{PoiseuilleEquation}
	\eta\left(\partial_x^2+\partial_y^2\right)u+\kappa\partial_z^2 u-\partial_z P
	\nonumber\\
	&-\frac{|e|B}{6c}\frac{v_Fu}{\left(v_F^2-u^2\right)^2}\left(\frac{2}{\tau}+\frac{3v_F^2+u^2}{v_F^2-u^2}\partial_zu
	\right.
	\nonumber\\
	&\left.
	+\frac{3v_F^2u-u^3}{v_F^2}\frac{\partial_z T}{T}\right)T^2=0.
\end{align}
At small velocities $u$ and temperatures $T$, the term $\kappa \partial_z^2 u$  and contributions
 in the last two lines of Eq.~\eqref{PoiseuilleEquation}, of the order of $uT^2$ in temperature
and velocity, can be neglected. Equation~\eqref{PoiseuilleEquation} then matches the equation for the flow of a
conventional liquid in a pipe~\cite{Landafshitz6,Suterra:PioseuilleReview}.

In accordance with the Hagen–Poiseuille equation~\cite{Landafshitz6,Suterra:PioseuilleReview},
the hydrodynamic velocity $u$ of such a liquid in the middle of the junction is given by
\begin{align}
	u=\frac{\zeta S}{\eta L}\Delta P,
\end{align}
where $\zeta$ is a coefficient of order unity that depends on the transverse shape of the junction; $L$ is the length
of the junction; $S$ is its cross-sectional area and $\Delta P$ is the pressure difference between the two ends of the junction.

The pressure difference $\Delta P$ may be generated by different temperatures at the ends of the junction.
Utilising Eq.~\eqref{PressureMain}, we estimate the flow velocity of the Weyl liquid as
\begin{align}
	u \sim \frac{|e|B}{c}\frac{S}{M}\frac{\partial_z T}{T},
	\label{uEstimate}
\end{align} 
where the ``mass'' $M$ describes the inverse curvature of the quasiparticle dispersion and is introduced in Eq.~\eqref{DispersionNonLinear}.
Using Eq.~\eqref{uEstimate} and assuming that the energy scale $Mv_F^2$ is given by the quasiparticle bandwidth and is of the order of $1 eV$,
we estimate that velocities $u$ of the order of $v_F\sim 10^8\frac{cm}{s}$ can be achieved 
in a junction of size $\sqrt{S}\sim L\sim100 nm$ (in all dimensions)
in a magnetic field $B\sim 1T$ and for temperature gradients $\partial_z T
\sim T/L$. The hydrodynamic regime is further favoured by larger system sizes and magnetic fields.


The flow of the electron liquid is associated with the heat flux (energy current) in the system, given by
\begin{align}
	q&=
	\frac{|e|B}{2\pi c}v_F\int\frac{dk}{2\pi}v_F k\left[f_L(k,u,\mu_L)-f_L(k,0,0)\right]
	\nonumber\\
	&-\frac{|e|B}{2\pi c}v_F\int\frac{dk}{2\pi}(-v_F k)\left[f_R(k,u,\mu_R)-f_R(k,0,0)\right]
	\nonumber\\
	=& \frac{|e|B v_F^2}{8\pi^2 c}
	\left[\frac{\mu_L^2}{(v_F-u)^2}-\frac{\mu_R^2}{(v_F+u)^2}\right]
	+\frac{|e|B}{6c}\frac{v_F^3 u}{(v_F^2-u^2)^2}T^2,
	\label{EnergyCurrent}
\end{align}
which can be used to detect the hydrodynamic flow and measure the average velocity $u$ of the flow.

In the absence of the electric field $\bE$, there is no electric current flowing through the system,
as follows from Eqs.~\ref{CNL} and \ref{CNR} and the charge neutrality condition $N_L+N_R=const$,
which require $\frac{\mu_L}{v_F-u}=\frac{\mu_R}{v_F+u}$. According to Eq.~\eqref{EnergyCurrent}, the energy current in the absence of the charge current is, therefore,
proportional to the hydrodynamic velocity $u$ of the current:
\begin{align}
	q\approx\frac{|e|B}{6c}\frac{u}{v_F}T^2\sim \left(\frac{|e|B}{c}\right)^2\frac{S}{M v_F}T\partial_z T.
\end{align}
The hydrodynamic flow can, thus, be generated by a temperature difference
$\Delta T$ at the ends of the junction and detected through the temperature-
and magnetic-field dependence of the heat conductance $G_q(T,B)$,
the response of the total energy flux to $\Delta T$.
Estimating the gradient $\partial_z T$ of the temperature as 
$\partial_z T\sim \Delta T/L$, where $L$ is the length of the junction, gives
\begin{align}
	G_q\sim \left(\frac{|e|B}{c}\right)^2\frac{S}{LMv_F}T. 
\end{align}



\section{Conclusion}

\label{sec:Conclusion}

In conclusion, we have studied the hydrodynamic motion of the electron liquid in a Weyl semimetal with two Weyl nodes in a strong magnetic field.
Such systems provide a conveniently accessible platform for achieving the hydrodynamic regime of transport because the impurity scattering rate
of Weyl fermions is strongly suppressed for certain directions of the magnetic field, perpendicular to the separation of Weyl nodes in momentum space.

Because Weyl fermions in a quantising magnetic field move parallel or antiparallel to the field, the motion of the liquid
resembles Poiseuille flow of a conventional liquid in a pipe (see Fig.~\ref{fig:flowpicture}).
The viscosity of such a liquid is dominated by the interactions between parallel layers of the liquid moving with
different velocities.
We have derived the hydrodynamic equations of motion of such a liquid for a Weyl semimetal with two Weyl nodes 
and computed microscopically its viscosity.
For realistic temperatures, the temperature dependence of the viscosity is given by $\eta(T)\propto T^2$.
The hydrodynamic flow of the electron liquid in a Weyl-semimetal junction
can be generated by a temperature gradient 
and probed via the heat conductance $G_q\propto B^2T$ of the junction.\\

We are grateful to I.~Shovkovy for numerous insightful discussions.
Part of this work was performed at the Aspen Center for Physics, which is supported by the National Science Foundation grant PHY-1607611.

\onecolumngrid
\vspace{2cm}
\cleardoublepage

\appendix

\section{Estimates of scattering rates in a strong magnetic field}
\label{sec:EstimatesScatteringRate}

In this section, we provide estimates of the internodal elastic scattering rate~\cite{Bednik:WeylMagnetotransport}
\begin{align}
	\label{ImpScatteringRate}
	\frac{1}{\tau} &\approx \frac{2\pi n_\text{imp}|e|Bv_F}{c\hbar}\left(\frac{e^2}{\hbar\epsilon v_F} \frac{\cos\theta}{2Q^2}\right)^2
\end{align}
in a Weyl semimetal in the ultraquantum regime, where $2Q$ is the momentum separation between the nodes; $n_\text{imp}$
is the concentration of impurities; $\epsilon$ 
is the dielectric constant
and $\theta$ is the angle between the direction of the field and the separation of the nodes. 
Because the scattering rate strongly depends on the node separation $2Q$, $1/\tau\propto Q^{-4}$,
it is rather sensitive to the details of the band structure and may differ by orders of magnitude 
even in Weyl semimetals with close parameters.

\begin{table}[H] 
	\centering
		\begin{tabular}{|c|c|c|c|c|}
			\hline
			Compound & $v_F \times 10^7 cm/s $ & $Q\times10^6 cm^{-1}$ & Source for $Q$
			and $v_F$ & $\frac{1}{\tau}\times 10^7 \cos^2{\theta} s^{-1}$\\
			\hline
			TaAs & 3.2 & 1.425 & ~\cite{grassano2018validity} & 9.71 \\
			\cline{1-3} \cline{5-5}
			NbAs & 3.0 & 0.473 & &853\\
			\hline
			WTeS & $4.0$  & 37.8 &~\cite{meng2019type:WTeS}\footnote{To our knowledge, there is no available
				data for the Fermi velocity in these compounds, which is why 
				we use the value $v_F=4.0\cdot 10^7 cm/s$.
			} & $1.57 \times 10^{-5}$  \\
			\cline{1-3} \cline{5-5}
			WTeSe & $4.0$  &26.2 &  &$6.80 \times 10^{-5}$  \\
			\hline
		\end{tabular}
		\caption{
			\label{scattering table}
			The Fermi velocities, node separation and scattering rates \eqref{ImpScatteringRate}
			for four type-I Weyl semimetals in the magnetic 
			field $B=1T$.
			For the estimates, 
			we used $n=10^{15}cm^{-3}$ and $\epsilon=30$.}
\end{table}

In Table.~\ref{scattering table}, we provide scattering rates for several of the Weyl semimetals.
These estimates show that even in materials with the strongest scattering, the elastic scattering length in the ultraquantum limit
 may be on the order of $10^{-2} cm$ 
and may exceed or be comparable to the size of the sample even for angles $\theta$ away from $\pi/2$, which makes the hydrodynamic regime conveniently accessible.


\section{Hydrodynamic pressure of the Weyl liquid}

\label{sec:Pressure}

The Sommerfeld expansion of the grand potential of the electron liquid in volume $V$
\begin{align}
	\Omega(T) = -TV\int d\varepsilon\, {\cal N}_0(\varepsilon)\ln\left(1+e^{\frac{\mu-\varepsilon}{T}}\right)=
	-\int d\varepsilon \frac{N(\varepsilon)}{e^{\frac{\varepsilon-\mu}{T}}+1},
\end{align}
gives
\begin{align}
	\Omega(T)=\Omega(0)-\frac{\pi^2}{6}V{\cal N}_0(\mu)T^2,
	\label{Omega}
\end{align}
where ${\cal N}_0(\varepsilon)$ is the density of states and $N(\varepsilon)$ is the number of electron states in the system 
with energies smaller than $\varepsilon$. Near the nodes of a Weyl semimetal in a strong magnetic field, the density of states per node is given by
${\cal N}_0(\varepsilon)=\frac{|e|B}{4\pi^2 cv_F}$.

Using that $\Omega=-PV$, we obtain the pressure of the equilibrium electron gas in a two-node Weyl semimetal in the form
\begin{align}
	P=P_0+\frac{|e|B}{12 cv_F}T^2,
\end{align}
where $P_0$ is a temperature-independent contribution which depends on the details of the quasiparticle dispersion away from the Weyl nodes.


\section{Details of the calculation of the polarisation operator}

\label{sec:PolarisationOperator}

In this section, we provide the details of the calculation of the polarisation operator in a Weyl liquid in a magnetic field in the ultraquantum limit,
in which only the zeroth Landau level contributes. In what follows, we use the Landau gauge
\begin{align}
	\bA=\left(-By,0,0\right)
	\label{Gauge}
\end{align}
for the vector potential of the magnetic field. For this gauge, the momentum $\bk_{xz}$ in the $xz$ plane is a good quantum number.

To the lowest order in interactions, the Matsubara polarisation operator is given by
\begin{align}
	\Pi(\br,\br^\prime,i\Omega)=2T\sum_{i\omega}\sum_{\bk_{xz},\bk_{xz}^\prime}
	\frac{\Psi_{\bk_{xz}}^*(\br)\Psi_{\bk_{xz}}(\br^\prime)\Psi_{\bk_{xz}^\prime}^*(\br^\prime)\Psi_{\bk_{xz}^\prime}(\br)}
	{\left(i\omega+i\Omega-E_{\bk_{xz}}\right)\left(i\omega-E_{\bk_{xz}^\prime}\right)},
	\label{PolOperatorGen}
\end{align}
where a prefactor of $2$ accounts from the presence of two nodes in the Weyl liquid, which contribute equally to the polarisation operator;
$E_{\bk_{xz}}$ is the quasiparticle dispersion at the zeroth Landau level with the momentum $\bk_{xz}$ in the $xz$ plane and
\begin{align}
	\Psi_{\bk_{xz}}(\br)=H_{\bk_{xz}}(y)\exp\left[{i(k_x x+k_z z)}\right]
\end{align}
is the orbital part of the corresponding wavefunction, where 
\begin{align}
	H_{\bk_{xz}}(y)
	={\left(\pi l_B^2S_{xz}^2\right)^{-\frac{1}{4}}}
	\exp
	\left[-\frac{1}{2}\left(y/l_B-k_x l_B\right)^2\right];
	\label{Hwf}
\end{align}
$S_{xz}$ is the area of the $xz$ cross-section of the system, which, for simplicity, is assumed to be constant along the $z$ axis;
$l_B$ is the magnetic length given by Eq.~\eqref{MagneticLength}.

For the evaluation of the viscosity in this paper, we focus on the correlations of electron densities on distances
$|\br-\br^\prime|\gg l_B$ significantly exceeding the magnetic length $l_B$. To evaluate the polarisation operator~\eqref{PolOperatorGen} at these scales,
 we first Fourier-transform it
with respect to the coordinate differences $x-x^\prime$ and $z-z^\prime$, using the exact translational invariance along the $x$ and $z$ axes:
\begin{align}
	\Pi(\bq_{xz},y,y^\prime,i\Omega)=2T\sum_{i\omega}\sum_{\bk_{xz}}
	\frac{H_{(\bk+\bq)_{xz}}^*(y)H_{(\bk+\bq)_{xz}}(y^\prime)H_{\bk_{xz}}^*(y^\prime)H_{\bk_{xz}}(y)}
	{\left[i\omega+i\Omega-E_{(\bk+\bq)_{xz}}\right]\left(i\omega-E_{\bk_{xz}}\right)}.
	\label{PolOperatorSimplified}
\end{align}

For long distances, $|\br-\br^\prime|\gg l_B$, it is sufficient to consider only small momenta $|\bq_{xz}|\ll l_B^{-1}$.
Because the characteristic lengthscale of the function $H_\bk(y)$, given by Eq.~\eqref{Hwf} is $l_B$, it allows us to neglect the momentum $\bq_{xz}$
in Eq.~\eqref{PolOperatorSimplified} and make the approximations $H_{(\bk+\bq)_{xz}}^*(y)\approx H_{\bk_{xz}}^*(y)$,
$H_{(\bk+\bq)_{xz}}(y^\prime)\approx H_{\bk_{xz}}(y^\prime)$.

The summand in Eq.~\eqref{PolOperatorSimplified} is peaked at the values of $y$ and $y^\prime$ given by $k_xl_b^2$ and has
a characteristic width of $l_B$ with respect to both of these coordinates. Taking into account the summation with respect to all values of $k_x$,
the operator may be considered, at distances $|y-y^\prime|\gg l_B$, as a sharply peaked function of $y-y^\prime$ and approximated
as
\begin{align}
	\Pi(\bq_{xz},y,y^\prime,i\Omega)
	=&
	\delta(y-y^\prime)\int\left\{
	2T\sum_{i\omega}\sum_{\bk_{xz}}
	\frac{H_{\bk_{xz}}^*(y)H_{\bk_{xz}}(y^\prime)H_{\bk_{xz}}^*(y^\prime)H_{\bk_{xz}}(y)}
	{\left[i\omega+i\Omega-E_{(\bk+\bq)_{xz}}\right]\left(i\omega-E_{\bk_{xz}}\right)}\right\}dy^\prime
	\nonumber\\
	=&2\delta(y-y^\prime)T\sum_{i\omega}\sum_{\bk_{xz}}
	\frac{H_{\bk_{xz}}^*(y) H_{\bk_{xz}}(y)}
	{\left[i\omega+i\Omega-E_{(\bk+\bq)_{xz}}\right]\left(i\omega-E_{\bk_{xz}}\right)}
	\nonumber\\
	=&2\delta(y-y^\prime)T\sum_{i\omega}\int \frac{dk_z}{2\pi}\frac{dk_x}{2\pi}
	\frac{1}{l_B\sqrt{\pi}}
	\frac{\exp\left[-(l_B k_x-y/l_B)^2\right]}
	{\left[i\omega+i\Omega-E_{(\bk+\bq)_{xz}}\right]\left(i\omega-E_{\bk_{xz}}\right)}
	\nonumber\\
	=&\frac{\delta(y-y^\prime)}{2\pi l_B^2}\sum_{i\omega}\int\frac{dk_z}{2\pi}
	\frac{1}{\left[i\omega+i\Omega-E_{(\bk+\bq)_{xz}}\right]\left(i\omega-E_{\bk_{xz}}\right)},	
	\label{PolOperator1}
\end{align}
where we have taken into account the dispersion $E_\bk$ depends only on the momentum component $k_z$ along the magnetic field
and is independent of the component $k_x$. 

Fourier-transforming Eq.~\eqref{PolOperator1} gives
\begin{align}
	\Pi(\bq,i\Omega)=
	\frac{|e|B}{2\pi c}\sum_{i\omega}\int\frac{dk_z}{2\pi}\frac{1}{\left[i\omega+i\Omega-E_{(\bk+\bq)_z}\right]\left(i\omega-E_{k_z}\right)},
	\label{P1}
\end{align}
which matches the polarisation operator of an effectively one-dimensional systems with the dispersion $E_{k_z}$ and a degeneracy
of $\frac{|e|B}{2\pi c}$ per transverse area.
The analytic continuation of Eq.~\eqref{P1} from the upper half-plane of Matsubara frequencies,
$i\Omega\rightarrow\varepsilon+i0$, to real frequencies $\varepsilon$ gives the retarded polarisation operator~\eqref{PolarisationOperatorWorking}.

\twocolumngrid

%


\end{document}